\documentstyle[aps,epsf,preprint,graphicx,prl]{revtex}

\begin{document} 
\draft

\def\beq{\begin{equation}}
\def\eeq{\end{equation}}
\def\beqn{\begin{eqnarray}}
\def\eeqn{\end{eqnarray}}
\def\btimes {\mbox{\boldmath $\times$}}
\def\bbox {\mbox{\boldmath $\box$}}
\def\bvarphi {\mbox{\boldmath $\varphi$}}
\def\ed{\end{document}}

\def\veps {{\varepsilon}}
\def\I {{\bf I}}
\def\II {{\bf II}}
\def\III {{\bf III}}
\def\IV {{\bf IV}}
\def\V {{\bf V}}
\def\VI {{\bf VI}}
\def\J {{\bf J}}
\def\H {{\bf H}}
\def\E {{\bf E}}
\def\1 {{\bf 1}}
\def\2 {{\bf 2}}
\def\3 {{\bf 3}}
\def\P {{\bf P}}
\def\r {{\bf r}}
\def\k {{\bf k}}
\def\p {{\bf p}}
\def\n {{\bf n}}
\def\A {{\bf A}}
\def\bv {{\bf v}} 
\def\AAN {$\!\!\!$ A$^{^{\!\!\!\!\! {\tiny {\circ}}}}$}
\def\aaN {$\!\!$ a$^{^{\!\!\!\! {\tiny {\circ}}}}$}

\title{On interacting fermions and bosons with definite total momentum}

\author{Ofir E. Alon\footnote{E-mail: ofir@tc.pci.uni-heidelberg.de}, Alexej I. Streltsov,
and Lorenz S. Cederbaum}
\address{Theoretische Chemie, Physikalisch-Chemisches Institut, Universit\"at Heidelberg,
Im Neuenheimer Feld 229, D-69120 Heidelberg, Germany}

\maketitle

\begin{abstract}

Any {\it exact} eigenstate with a definite momentum of a many-body Hamiltonian can be written
as an integral over a {\it symmetry-broken} function $\Phi$.
For two particles, we solve the problem {\it exactly} 
for all energy levels and any inter-particle interaction.
Especially for the ground-state, $\Phi$ is given by 
the simple Hartree-Fock/Hartree ansatz for fermions/bosons.
Implications for several and many particles as well as a numerical example are provided.
\end{abstract}

The homogeneous gas of interacting fermions and bosons is a fundamental concept in the physics of
many-particle systems, 
see e.g., Refs.~\cite{Fetter_book,Ripka_book}.
Expect for specific cases, 
such a many- or several-body problem cannot be solved exactly,
i.e. the wavefunctions and energies of the ground and excited states are not known. 
In fact, for more than a few particles it becomes already impossible to numerically compute the 
exact ground-state energy and wavefunction.
Consequently, approximations are a must.

In this Letter, we introduce an exact continuous configuration-interaction ansatz for
the many-body wavefunction $\Psi$ of interacting particles in a volume with periodic boundary conditions.
Examples for realizations of this case are a ring in one dimension (1D),
a torus, a long, thin pipe (tube) or a sphere in 2D,
and the text-book example of a ``big box'' in 3D.
Specifically, we employ a many-body function $\Phi$
as a basis function for a continuous expansion of $\Psi$.
The shape of $\Phi$ is to be optimized by employing the variational principle.
For two particles, we solve the problem {\it exactly} 
for {\it all} energy levels and {\it any} inter-particle interaction.
Especially for the exact ground-state, $\Phi$ is given by 
the simple Hartree-Fock/Hartree ansatz for fermions/bosons.
For more particles as will be explained below,
our ansatz with any specific choice for $\Phi$ 
is better than solving the Schr\"odinger equation by optimizing $\Psi=\Phi$ itself.
For instance, taking for $\Phi$ a Hartree-Fock/Hartree ansatz, the resulting equations
would lead to lower energies than the corresponding Hartree-Fock/Hartree equations.   
These properties make our ansatz a particularly attractive approximation 
for the few as well as the many-body problem.

Consider the {\it generic} many-body Hamiltonian describing $N$ fermions or bosons in a 3D box of volume $V=L_x L_y L_z$: 
\beq\label{Ham_MB}
 \hat H(\r_1,\r_2,\ldots) = -\sum_{i=1}^{N} \frac{\partial^2}{\partial \r_i^2} + 
           \sum_{i>j=1}^N U(\r_i-\r_j),
\eeq
Here, $\r_i=(x_i,y_i,z_i)$ (we use Cartesian coordinates) is the coordinate the $i$-th particle and $U(\r_i-\r_j)$ 
describes a {\it general} pairwise interaction between the $i$-th and $j$-th particles.
Periodic boundary conditions in 3D are assumed.

The basic question we would like to address here is how to approach the solutions of Eq.~(\ref{Ham_MB}).
Since {\it continuous} translational symmetry exists in each dimension, 
we construct a continuous {\it self-consistent} configuration-interaction (CCI) ansatz 
for the many-body wavefunction.
Namely, we define the CCI wavefunction as
\beq\label{CCI_def}
 \Psi(\r_1,\r_2,\ldots) = \int d\r_0 \, C(\r_0) \, \Phi (\r_1-\r_0,\r_2-\r_0,\ldots),
\eeq
where {\it both} the weight $C(\r_0)$ and the many-body function $\Phi(\r_1,\r_2,\ldots)$ are to be determined 
from the requirement that the variation of the energy per particle  
$\varepsilon(N) = \frac{1}{N}\frac{<\Psi|\hat H|\Psi>}{<\Psi|\Psi>}$ vanishes. 
$\r_0$ is a vector in 3D.
Demanding stationarity of $\varepsilon(N)$ with respect to $C(\r_0)$, 
we proved that
any variational solution $\Psi$ must be an eigenfunction of the momentum, say with the momentum $\P_0$,
where $\P_0=2\pi\left(\frac{n_x}{L_x},\frac{n_y}{L_y},\frac{n_z}{L_z}\right)$ and the $n_i$ take some specific integer values.
This can most generally be fulfilled by 
fixing $C(\r_0) = e^{+i \P_0 \cdot \r_0}$ and, thereby, allowing 
 $\Phi$ to be a {\it symmetry-broken} function 
(we use this term to describe a function which is not an eigenfunction of the momentum).
The decomposition of $\Psi$ in terms of a {\it symmetry-broken} $\Phi$ 
provides many more degrees of freedom for optimization (this will be evident below), 
and will lead to lower energies and better approximations for the wavefunctions.
In fact, any {\it exact} solution of Eq.~(\ref{Ham_MB})
 with the momentum $\P_0$ can be decomposed into a {\it symmetry-broken} function $\Phi$
(such a decomposition is not unique). 
This finding is central to this work.
The proof of this property is straightforward to make.
The integration over $\r_0$ eliminates from $\Phi(\r_1-\r_0,\r_2-\r_0,\ldots)$
any component which is not an eigenfunction of the momentum with the eigenvalue $\P_0$.
Of course, $\Psi$ should possess the other symmetries of the many-body problem (\ref{Ham_MB}).
Specifically, it should be antisymmetric or symmetric 
under permutations of any two particles for fermions or bosons, respectively, and be an eigenfunction of the total spin.

For two particles we can explicitly determine $\Phi$ which reproduces the
{\it exact} solution of the Schr\"odinger equation for all energy levels and any interaction $U$.
Specifically, we will prove below that taking for $\Phi$ in Eq.~(\ref{CCI_def}) an Hartree-Fock/Hartree ansatz
for two fermions/bosons is exact for all eigenstates of (\ref{Ham_MB}) with $\P_0=0$, 
among which is the ground state. 
The case of $\P_0\ne 0$ is also exact with another ``simple'' form of $\Phi$ and
will be presented thereafter.
We refer in the following to the spatial degrees of freedom alone also for two fermions 
since the spin and spatial parts of the wavefunction are separable in this case. 
A two-particle (spatial) wavefunction symmetric under permutations of particles' indices, $\Psi_+(\r_1,\r_2)$, 
describes bosons and also fermions in the singlet state, 
whereas an anti-symmetric wavefunction, $\Psi_-(\r_1,\r_2)$, 
describes fermions in the triplet state. 
Accordingly, the general solution (with $\P_0=0$) takes on a simple form
\beqn\label{solution_L0}
\Psi_+(\r_1,\r_2) &=& \frac{1}{2} \sum_{\k,k_x\ge0} A_\k \cos[\k \cdot (\r_1-\r_2)], \nonumber \\ 
\Psi_-(\r_1,\r_2) &=& \sum_{\k,k_x>0} A_\k \sin[\k \cdot (\r_1-\r_2)] \
\eeqn
where $\k=2\pi\left(\frac{n_x}{L_x},\frac{n_y}{L_y},\frac{n_z}{L_z}\right)$ is the momentum 
and the $n_i$ take integer values.
The coefficients $A_\k$ are {\it real} 
and depend on the nature of the two-body interaction in Eq.~(\ref{Ham_MB})
and, of course, on the energy level and particle statistics.
Next, we introduce the following two orthogonal orbitals:
\beqn\label{orbital_L0}
 \phi_1(\r) = \frac{1}{\sqrt{V}} \sum_{\k,k_x\ge0} \sqrt{A_\k} \cos(\k \cdot \r), \nonumber \\ 
 \phi_2(\r) = \frac{1}{\sqrt{V}} \sum_{\k,k_x>0} \sqrt{A_\k} \sin(\k \cdot \r). \
\eeqn
After some algebra, it is straightforward to show that with these orbitals the CCI ansatz (\ref{CCI_def}) is {\it exact},
namely that the following decompositions with Hartree and Hartree-Fock wavefunctions exist:
\beqn\label{decomposition_L0}
 \Psi_+(\r_1,\r_2) &=& \int d\r_0 
 \phi_1(\r_1-\r_0) \phi_1(\r_2-\r_0), \nonumber \\
 \Psi_-(\r_1,\r_2) &=& \int d\r_0 
\det| \phi_1(\r_1-\r_0) \phi_2(\r_2-\r_0)|. \
\eeqn

Let us discuss the results just obtained for $\P_0=0$.
The CCI orbitals (\ref{orbital_L0}) are very unique.
They do not possess the translational symmetry of the problem, i.e., they are {\it symmetry-broken} functions,
even for very weakly-interacting particles, for which the Hartree and Hartree-Fock orbitals are 
 symmetry preserving \cite{Fetter_book,Ripka_book}.
They are, however, eigenfunctions of the space inversion operator, $\r \to - \r$.
Another interesting property is that $\phi_1(\r)$ and $\phi_2(\r)$ are, in general, complex quantities.
Nonetheless, the resulting {\it exact} wavefunction is real and possesses the correct translational symmetry.
In other words, these additional degrees of freedom the CCI ansatz allows for its orbitals
make it {\it exact} with an Hartree-Fock/Hartree ansatz for $\Phi(\r_1,\r_2)$,
no matter what the inter-particle interaction is. 
Notwithstanding, applying these mean-fields themselves for two particles only is an inaccurate approximation,
to say least.

Next, we turn to lay out the case of $\P_0\ne 0$.
Here, the wavefunction may be written as
\beq\label{solution_L_ne_0}
\Psi_\pm(\r_1,\r_2) = \sum_\k A_\k \left[ 
 e^{+i \k \cdot (\r_1-\r_2)} \times e^{+i \P_0 \cdot \r_1} \pm 
 e^{-i \k \cdot (\r_1-\r_2)} \times e^{+i \P_0 \cdot \r_2} \right]
\eeq
($A_\k$ are complex coefficients). 
Introducing the two general orbitals 
\beqn\label{orbital_L_ne_0}
 \phi_1(\r) &=&  \frac{1}{\sqrt{V}} \sum_{\k} \sqrt{A_\k} e^{i \k \cdot \r}, \nonumber \\
 \phi_2(\r) &=&  \frac{1}{\sqrt{V}} \sum_{\k} \sqrt{A_{\P_0-\k}} e^{i \k \cdot \r}, \
\eeqn
the CCI ansatz is again exact with the following decomposition
\beq\label{decomposition_L_ne_0}
 \Psi_\pm(\r_1,\r_2) = \int d\r_0 e^{+i \P_0 \cdot \r_0}
\left[\phi_1(\r_1-\r_0) \phi_2(\r_2-\r_0) \pm \phi_2(\r_1-\r_0) \phi_1(\r_2-\r_0) \right].
\eeq
This concludes our proof that the CCI ansatz with a ``simple'' choice for $\Phi(\r_1,\r_2)$ is exact for all two-particle states, 
for any inter-particle interaction and for fermion and boson statistics.

Having solved the problem for $N=2$, we turn to the cases of several and many particles.
Here we do not expect that an Hartree-Fock/Hartree ansatz for $\Phi$ will solve the problem exactly.
That there is such a $\Phi$ for any solution of (\ref{Ham_MB}) was proved above.
So, what can be said about an approximate $\Phi$?
For {\it any} ansatz for $\Phi$ 
the variational principle ensures that the ground-state energy obtained 
by {\it optimizing} $\Phi$ within the CCI ansatz 
is at least that obtained by utilizing $\Phi$ itself as the ansatz for $\Psi$.
In practice, the CCI energy will be {\it lower} (as found above for $N=2$),
since the CCI ansatz has many more degrees of freedom for optimizing $\Phi$ and minimizing the energy,
also see example below.
For instance, taking the Hartree-Fock/Hartree ansatz for $\Phi$ 
the CCI ansatz will provide lower energies than the Hartree-Fock/Hartree equations.
In conjunction with that, 
the CCI orbitals will be {\it symmetry-broken}, even for very weakly-interacting particles,
a situation found above for the exact solution of the two-particle problem, see Eq.~(\ref{orbital_L0}).
Contrast this with the Hartree-Fock/Hartree orbitals which are symmetry preserving for
not too strongly interacting particles \cite{Fetter_book,Ripka_book}.
All the above make the CCI ansatz particularly attractive to apply 
both for many- as well as for several-particle systems already with a Hartree-Fock/Hartree ansatz for $\Phi$.
Of course, other ansatzs for $\Phi$ can be plugged into Eq.~(\ref{CCI_def}), if needed. 

In the following, we would like to implement the CCI ansatz (\ref{CCI_def}) in its simplest form.
We concentrate on bosons, set $\Phi$ to be an Hartree product, $\Phi(\r_1,\r_2,\ldots) = \prod_{i=1}^N \, \phi(\r_i)$,
put $C(\r_0)=1$ for the ground state,
and derive the CCI equation for the (single-particle) orbital $\phi(\r)$.
For the inter-particle interaction in (\ref{Ham_MB}) we take the contact interaction $U(\r_i-\r_j) = U_0 \delta(\r_i-\r_j)$.
This interaction is widely used for the dilute Bose gas, see, e.g., Refs.~\cite{review1,review2} and references therein, 
where $U_0$ is proportional to the bosons' s-wave scattering length.

The final result for $\varepsilon(N)$ takes on the following appealing form:
$$
 \varepsilon(N) = \frac{\int d\r_0  S^{N-2}(\r_0) \varepsilon(N;\r_0)}
{\int d\r_0 S^N(\r_0)}, \ \ \  
S(\r_0) \equiv \left<\phi(\r)\left|\right.\phi(\r-\r_0)\right>,     \eqno{(9a)}
$$
with the energy density ($\hat T = - \frac{\partial^2}{\partial \r^2}$)
$$ 
 \varepsilon(N;\r_0) = \left[ S(\r_0) \cdot \left<\phi(\r)\left|\hat T\right|\phi(\r-\r_0)\right>
   + \frac{U_0(N-1)}{2} \left<\phi^2(\r)\left|\right.\phi^2(\r-\r_0)\right> \right]. \eqno{(9b)} \
$$
The Gross-Pitaevskii (GP) energy \cite{review1,review2} can be seen as a private case of the CCI energy (9) 
if we were to evaluate the latter for $\r_0=0$ only. 

By minimizing $\varepsilon(N)$ in (9) with respect to the orbital $\phi(\r)$ we obtain
an equation for this orbital.
The final expression reads 
$$ 
\!\int\! d\r_0 S^{N-2}(\r_0)\! \left[S(\r_0) \cdot \hat T +
 (N-1) U_0 \phi^\ast(\r) \phi(\r-\r_0) \right] \phi(\r-\r_0) = \! 
\int\! d\r_0  S^{N-3}(\r_0) \mu(\r_0) \phi(\r-\r_0), \eqno{(10a)} 
$$
with the chemical potential density 
$$
 \mu(\r_0) = S^{2}(\r_0) N \varepsilon(N) - (N-1) \varepsilon(N-1;\r_0). \eqno{(10b)} 
$$
Eq.~(10), the CCI equation as we shall call it, is an integro-differential equation for the orbital $\phi(\r)$
in the case of $N$ interacting bosons, which has to be solved self-consistently.
The non-trivial dependence of the CCI equation on the number of particles $N$ is 
via powers of the overlap $S(\r_0)$. 
It comes from the CCI ansatz which allows the function $\Phi$ and, hence, the 
orbital $\phi(\r)$ to be translated throughout space.
We will see the effect of this dependence in a numerical example below.

Having derived the CCI equation Eq.~(10)
we would like to solve it for a specific example,
namely for the attractive Bose gas in 1D.
This system has attracted a renewed interest recently, see, e.g., Refs.~\cite{Reinhardt,Ueda_PRA,Our} and references therein.
As mentioned in the introduction, in 1D we may equivalently speak of bosons on a ring.
In this case, it is convenient to re-scale the coordinate $x$ 
and define the angle $\varphi=\frac{2\pi}{L_x} x \in [-\pi,\pi)$.
The effective particle-particle interaction becomes then $\tilde{U}_0 \delta(\varphi_1-\varphi_2), 
\ \tilde{U}_0=\frac{U_0 L_x}{2\pi}$ and $\hat T = - \frac{\partial^2}{\partial \varphi^2}$.
The properties of the system are described by a single dimensionless parameter $\gamma=\frac{\tilde{U}_0(N-1)}{2\pi}$ \cite{Ueda_PRA},
which is negative in the attractive case.

Let us briefly describe the properties of the GP solution, as far as they are needed here.
Here, for weakly-interacting bosons, i.e. for $|\gamma|\le 0.5$, the ground-state orbital is 
a constant equal to $1/\sqrt{2\pi}$ \cite{Reinhardt,Ueda_PRA}, also see Fig.~1A.
For a stronger interaction, i.e., for $|\gamma|>0.5$, the GP solution lowest in energy becomes localized 
and, hence, symmetry-broken \cite{Reinhardt,Ueda_PRA}, also see Fig.~1B.
In both regimes, the shape of the GP orbital {\it does not} depend on $N$ 
for a fixed value of $|\gamma|$ \cite{Ueda_PRA}.
In Ref.~\cite{Our} it has been shown that when this {\it fixed}, symmetry-broken GP orbital
is plugged into a CCI ansatz the energy obtained is much lower than the GP one,
 and that 
``with increasing number of particles and/or strength of inter-particle interaction,
 is even {\it lower} than that accessed by tractable diagonalization of the many-body Hamiltonian.''
In this context, Ref.~\cite{Our} is a special case of this work; 
There, a {\it symmetry-broken}
orbital was ``supplied'' by the mean-field solution itself.
The breakthrough which we present here is the 
successful {\it optimization} of the CCI orbital $\phi$ by solving Eq.~(10) for bosons on a ring.
To this end, we employed the discrete variable representation (DVR) method \cite{DVR}
and solved Eq.~(10) iteratively (self-consistently) till convergence. 
The numerical implementation was straightforward.
Note that, although the wavefunction strongly depends on $N$, 
the numerical effort does not.

Analyzing the results, we start with the case of $N=2$ bosons.
The Hartree (GP) ansatz itself is not a good approximation for such a small number of bosons.
Yet, it becomes exact inside the CCI ansatz (\ref{CCI_def}), see proof above and Eq.~(\ref{decomposition_L0}).
For this, we examine the unique shape of the CCI orbital $\phi$.
For attractive bosons, this orbital for the ground state is a real function
(for repulsive bosons this orbital is a complex function).
The shape of $\phi$ for $N=2$ is presented in Fig.~1 for the two representative values 
$|\gamma|=0.2$ and $|\gamma|=1.0$. The corresponding GP orbitals are depicted for comparison.
As anticipated on the basis of the exact analysis, $\phi$ is
symmetry-broken also for values of $|\gamma|$ for which the GP orbital is not.
For any value of $\gamma$ the CCI orbital is totally different from the GP one:
it is narrower and on top of that exhibits a central spike.
Interestingly, the structure of $\phi$ below and above $|\gamma|=0.5$ is different.
This shows that the ``phase transition'' occurring at $|\gamma|=0.5$ manifests itself
not only within the mean-field (GP) solution \cite{Reinhardt,Ueda_PRA}.
Rather, it also appears in the exact solution represented by the CCI orbital depicted in Fig.~1. 

Finally, we set on to study the shape of the CCI orbital for $N>2$, 
i.e, the evolution of $\phi$ with increasing particle number as obtained by solving Eq.~(10) in 1D.
Here we do not expect that employing the Hartree ansatz for $\Phi$ 
will provide the {\it exact} energy.
It will, however, provide a very good estimate for it,
lower, naturally, from the already low energy obtained 
with the fixed--unoptimized, symmetry-broken GP orbital \cite{Our}.
In Fig.~2 we present for $N=5,25,100,1000$ and $10000$ and $|\gamma|=0.2$ the CCI orbital,
which is, as expected, symmetry-broken. 
The dependence of $\phi$ on $N$ 
comes from the overlap $S$ whose powers appear in the CCI equation (10), but, of course, are absent in the GP equation.
As the number of bosons is increased, the shape of the CCI orbital is flattened and tends to the
GP solution (which is constant at $|\gamma|=0.2$ for any $N$ \cite{Ueda_PRA}, see Fig.~1A).
Nonetheless, we see that the spike characterizes the CCI orbital for any number of particles, 
just as it does for the exact solution for $N=2$, see Fig.~1.
The same would hold true for other values of $|\gamma|$ and is a signature 
of the correlation energy accounted for by the CCI ansatz, 
and of the contact interaction $U$.

In conclusion, within the CCI ansatz 
any eigenstate with a definite momentum of a many-body Hamiltonian 
is exactly written in terms of a {\it symmetry-broken} function $\Phi$.
For two interacting fermions/bosons, 
$\Phi$ has been determined exactly for all energy levels and any inter-particle interaction. 
For more particles and any given ansatz for $\Phi$, 
the resulting CCI equation provides lower
energies than the ansatz $\Psi=\Phi$ can.
All of these makes our CCI ansatz a very attractive tool to be utilized in the
many-body problem of an homogeneous, interacting fermion/bosons system, already with a Hartree-Fock/Hartree ansatz for $\Phi$.
We briefly remark that since the CCI ansatz is a variational one,
it can be reformulated in the non-homogeneous case as well.
 
\acknowledgments

\noindent
We thank Kaspar Sakmann for discussions.


\begin{figure}[ht] 
\includegraphics[width=10cm,angle=-90]{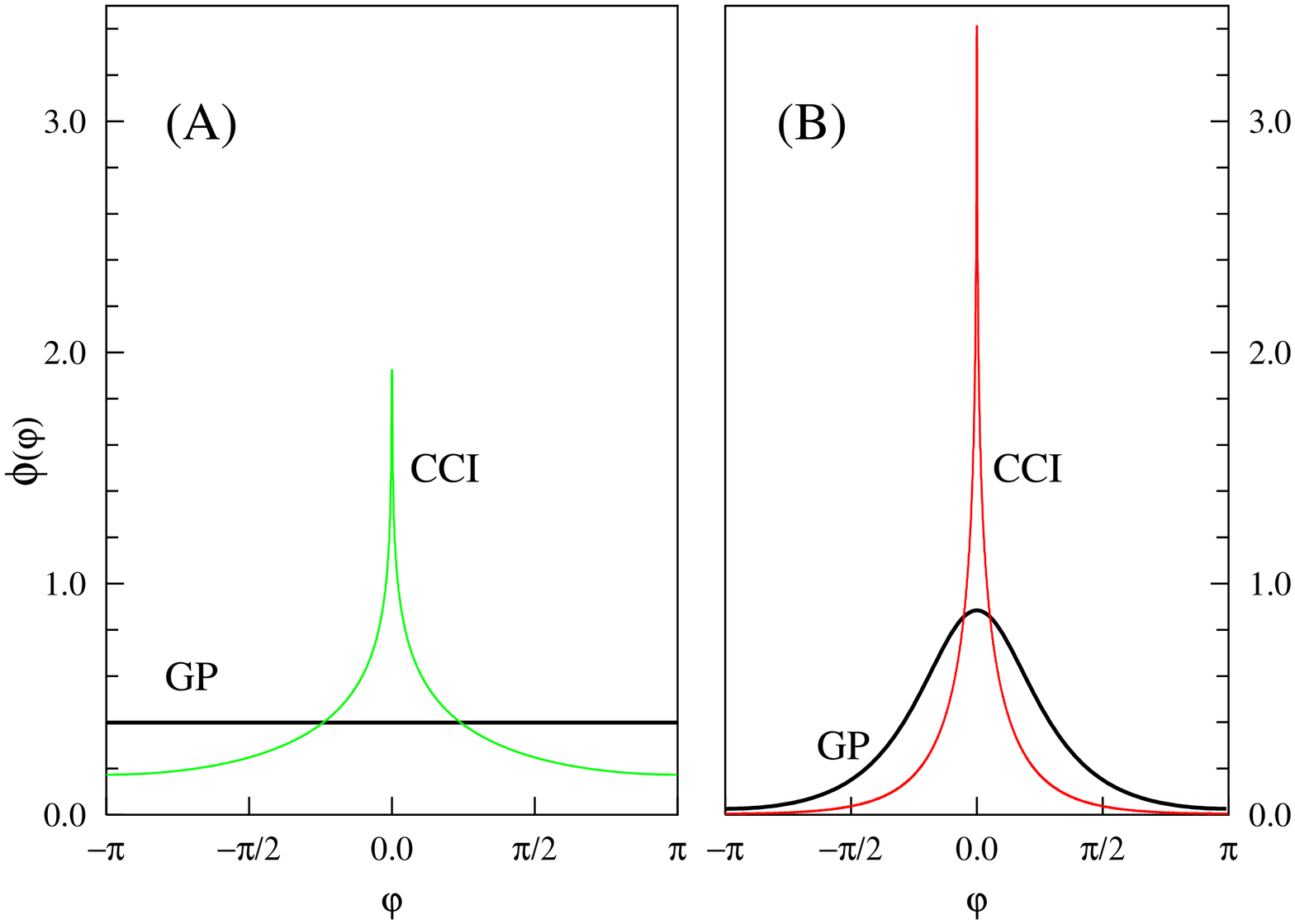}
\caption [kdv]{The CCI orbital which {\it exactly} solves the problem of $N=2$ attractive bosons on a ring.
For comparison, the corresponding GP orbital is presented: (A) $|\gamma|=0.2$; (B) $|\gamma|=1.0$.
}
\end{figure}


\begin{figure}[ht] 
\includegraphics[width=10cm,angle=0]{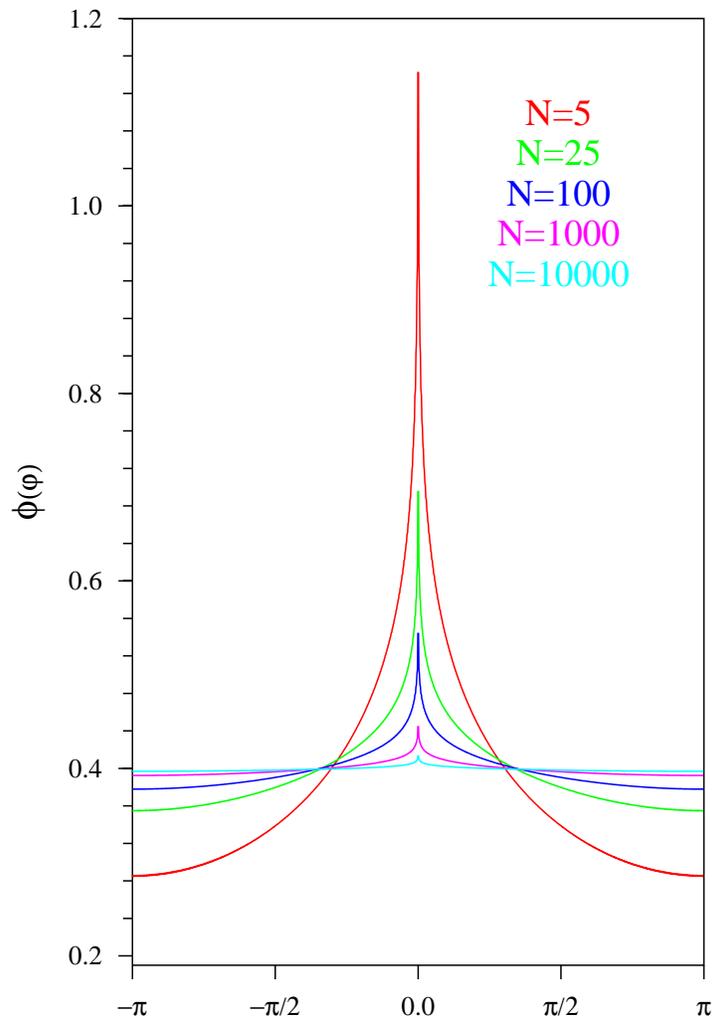}
\caption [kdv]{The CCI orbital for $N=5,25,100,1000$ and $10000$ attractive bosons on a ring
for $|\gamma|=0.2$.
}
\end{figure}

\ed